\documentclass[11pt,a4paper]{article}
\usepackage{jheppub}
\usepackage{bbold}
\usepackage{xcolor}
\newcommand{\del}{\partial}

\newcommand{\ud}{\mathrm{d}}
\newcommand{\V}{\mathcal{V}_E}

\newcommand{\tr}{\mathrm{tr}}
\newcommand{\Odd}{\textrm{O}(d,d)}
\newcommand{\figref}[1]{figure \ref{#1}}
\title{On de Sitter vacua in O(d,d) invariant cosmology}
\author[1,2]{Yang Liu, Antonio Padilla, Paul M. Saffin and Robert G. C. Smith}
 \affiliation[1]{School of Physics and Astronomy, University of Nottingham, University Park, Nottingham NG7 2RD, United Kingdom}
\affiliation[2]{The Nottingham Centre of Gravity, University of Nottingham, Nottingham NG7 2RD, UK}

\emailAdd{ppxyl1@exmail.nottingham.ac.uk}
\emailAdd{antonio.padilla@nottingham.ac.uk}
\emailAdd{paul.saffin@nottingham.ac.uk}
\emailAdd{ppxrs3@exmail.nottingham.ac.uk}
\abstract{We perform a thorough analysis of de Sitter solutions in $\Odd$ invariant cosmologies. Starting with a homogeneous and isotropic framework we examine conditions for the existence of such solutions to the vacuum field equations, non-perturbative in $\alpha'$ in both the string frame and the Einstein frame.  We elucidate the nature of the instability in the string frame vacuum. For the Einstein frame, we demonstrate that the de Sitter solutions cannot be eternal.  We then extend our analysis to include Bianchi I universes  where the $\Odd$ symmetry includes scale factor exchange as well as scale factor duality.  We show how the theory can be extended to the anisotropic case  so that it admits de Sitter solutions, noting the crucial role played by the $\Odd$ symmetry in satisfying any additional constraints.}
\begin{document}
\maketitle

\section{Introduction}
Cosmological observations \cite{cmb,SN1,SN2} indicate that our universe has undergone two phases of accelerated expansion, at early \cite{Baumann} and at late times \cite{Ed}.  At early times we have inflation, consistent with the evolution of a scalar field in slow roll.  At late times we have dark energy, which may correspond to our universe approaching a de Sitter vacuum, or else quintessence, which is a low scale version of early universe inflation. Data recently released by the DESI collaboration shows some preference for dark energy with an evolving equation of state \cite{DESI:2024mwx}, consistent with  quintessence. Whether it takes the form of quintessence or de Sitter, obtaining an accelerated expansion from string theory is an ongoing challenge.  Although we can construct stable de Sitter vacua using background flux and brane uplifts (see eg. \cite{KKLT,Tbranes} and for reviews \cite{Grana,STcos}) the consistency of these constructions is a matter of considerable debate (see eg. \cite{nods,dSVQ}).
This has led to the so-called de Sitter conjecture which implies that fully consistent and stabilised de Sitter vacua are absent from the spectrum of weakly coupled string theory at or near the boundary of moduli space \cite{ds1,ds2,ds3}. If the conjecture is correct and these vacua are indeed absent, then quintessence is also expected to be ruled out \cite{quin1,us1,us2,arthur}. 

String theory is a consistent quantum theory of gravity, with a well defined low energy description in terms of higher dimensional supergravities. For this to be relevant to our universe, it must be able to admit a compactification down to four dimensions, where gravity is well described by General Relativity and matter by a suitable extension of the Standard Model of Particle Physics. It should also be able to accommodate an accelerated cosmological expansion. Of course, it is still possible that the concerns surrounding de Sitter vacua in perturbative string theory will go away and we will be able to find consistent accelerated cosmological solutions that everyone is happy with. However, if this is not the case, we will need a non-perturbative understanding of the theory to probe the full vacuum structure and establish the existence of de Sitter vacua far from the boundary of moduli space. An important no-go theorem, valid at any order in the string scale, suggests that de Sitter vacua of dimension four or higher are not possible for the classical  heterotic string, or indeed classical type II strings in the absence of RR fluxes \cite{Sethi}. To say much more will likely require a third string revolution - a step change in our current approach to string theory/M theory. We can also consider the possibility that de Sitter is a resonant or excited state of string theory \cite{res1,res2,excitedstate}. 

We can try and get some insight into what can happen non-perturbatively by focussing on the effective field theory for classical strings in the homogeneous limit. The Neveu-Schwarz sector describing the metric, the Kalb-Ramond field and the dilaton, are known  to possess an $\Odd$ symmetry to leading order \cite{Veneziano1,Meissner1,Meissner2,Sen1,Gasperini:1991ak}. This symmetry is expected to be preserved to all orders in the slope parameter $\alpha'$, motivating corrections to the leading order Neveu-Schwarz theory using higher order $\Odd$ invariant operators \cite{Meissner3}. Note that $\Odd$ invariance can be made manifest using double field theory, albeit at the expense of manifest diffeomorphism invariance \cite{DFT,DFT2,Hohm:2013jaa,Hohm:2014xsa}. Duality invariant cosmology has been used to tackle the problem of the initial singularity (see eg.  \cite{Gasperini:2002bn,Wu,Brandenberger:2018xwl,Bran2018,Quintin}). 

Remarkably, for homogeneous fields, Hohm and Zwiebach have been able to provide a complete classicfication of all higher order terms  invariant under  $\Odd$ \cite{Hohm:2015doa,Hohm:2019ccp,Hohm:2019jgu}.  With field redefinitions, the system is reduced to first order time derivatives allowing them to write down the most general duality invariant theory to all orders in $\alpha'$ for the metric, Kalb-Ramond field and dilaton. To leading order in the expansion in $\alpha'$ the  theory
coincides with the Neveu-Schwarz sector of the supergravity theories describing the effective theory of classical strings, in the homogeneous limit. In the limit of vanishing Kalb-Ramond field and an isotropic spatially flat metric, the theory exhibits scale factor duality $a\to 1/a$ as a remnant of the underlying $\Odd$ symmetry.  The full theory is now tractable at all orders in $\alpha'$, enabling the authors of \cite{Hohm:2019ccp,Hohm:2019jgu} to identify string frame de Sitter solutions to the vacuum field equations that are {\it non-perturbative} in $\alpha'$, with a time dependent dilaton.  The dynamics of this new class of theories has been studied in a series of subsequent papers \cite{EFdS,Bernardo:2019bkz,Bernardo:2020zlc,Bernardo:2020nol,Rost, Bernardo:2021xtr,Quintin:2021eup,Bieniek:2022mrv,Bernardo:2022nex,Bieniek:2023ubx,Codina:2023fhy,Gasperini:2023tus,Conzinu:2023fth}. In particular, one can extend the system to include $\Odd$ invariant matter fields \cite{Gasperini:1991ak,Bernardo:2019bkz}, or find theories that admit de Sitter solutions even in Einstein frame \cite{EFdS,Bernardo:2019bkz, Rost}. Instabilities for homogeneous and isotropic perturbations were studied in detail in \cite{Bernardo:2020zlc,Rost,Bieniek:2022mrv,Bernardo:2022nex} with subsequent extensions to anisotropic perturbations \cite{Bieniek:2023ubx}. Bianchi I universes were studied in \cite{Codina:2023fhy,Gasperini:2023tus,Conzinu:2023fth}, along with two-dimensional black hole solutions \cite{Codina:2023fhy} and bouncing cosmologies \cite{Gasperini:2023tus,Conzinu:2023fth}.
 
In this paper, we revisit the dynamics of $\Odd$ invariant cosmologies, starting with isotropic and spatially flat metrics before moving on to the anisotropic case. Our focus is on the existence of de Sitter solutions in both the string frame and the Einstein frame, with a time dependent dilaton. Of course, the isotropic case is simpler and much more extensively studied. Within the isotropic framework, string frame de Sitter solutions are known to exist but have been shown to be unstable \cite{Bernardo:2020zlc,Bieniek:2022mrv,Bernardo:2022nex}, something we are able to verify using slightly different methods. de Sitter solutions in the Einstein frame have been obtained in the presence of a source \cite{Bernardo:2019bkz} and in vacuum \cite{EFdS, Rost}. The latter only exist when the theory takes on a very bespoke form at all orders in $\alpha'$. As we will show, these Einstein frame de Sitter solutions are the only solutions to the vacuum field equations for these particular theories for a finite time period,  ensuring their local stability under homogeneous fluctuations.  However, the solutions are not eternal in both past and future. This is because the dilaton evolves. After a finite time (towards either the past or future), the solution is kicked into a new dynamical phase where an Einstein frame de Sitter metric is no longer possible.  This resonates (!) with the idea of de Sitter as a resonance \cite{res1,res2}. 

These solutions enjoy a natural  extension to anisotropic universes thanks to the $\Odd$ symmetry. Here we consider a Bianchi I universe, with a number of different scale factors with corresponding Hubble parameters.  In a notable departure from the  isotropic framework, the $\Odd$ symmetry now implies an invariance under the exchange of any two of those Hubble parameters, placing important restrictions on the structure of the underlying theory.  Although isotropy is no longer assumed at the level of the theory, we are interested in isotropic solutions: namely, the de Sitter solutions in both string frame and Einstein frame, with a time dependent dilaton. In string frame, de Sitter solutions are possible, provided the anisotropic theory obeys properties analogous to the isotropic case. These solutions are once again unstable. In the Einstein frame, we find conditions for the de Sitter solutions to exist. This mirrors the isotropic case discussed in the previous paragraph, with additional conditions pointing along the anisotropic directions. Remarkably, these extra conditions hold automatically thanks to the $\Odd$ symmetry.

The rest of this paper is organized as follows. In section \ref{sec:review} we give a brief review of $O(d,d)$-invariant cosmology. In section \ref{sec:iso}, we review the structure of the underlying theory for an isotropic and spatially flat universe, including sources. In section \ref{sec:SF} we review the derivation and instability of de Sitter solutions in the string frame, providing some added insight into the source of the instability. In section \ref{sec:EF} we demonstrate how de Sitter solutions can also be found in the Einstein frame. As discussed above, these solutions exist for a finite time period, during which time they are stable against homogeneous fluctuations. However, the solutions are not external, and will eventually give way to a new cosmological phase. In section \ref{sec:aniso}, we change gears and extend our analysis to include Bianchi I universes. In section \ref{sec:SFan} we establish the appropriate conditions for string-frame de Sitter solutions, commenting on the stability. In section \ref{sec:EFan} we  do the same for de Sitter solutions in the Einstein frame, noting the crucical role played by the $\Odd$ symmetry. In section \ref{sec:conc}, we summarise our conclusions, and extend the results to non-critical dimensions. 

\section{Review of O$(d,d)$-invariant cosmology} \label{sec:review}
We begin with a review of all orders in $\alpha'$ duality invariant cosmology,  as originally proposed by Hohm and Zwiebach \cite{Hohm:2019ccp,Hohm:2019jgu} building on earlier pioneering work \cite{Veneziano1,Meissner1,Meissner2,Sen1,Gasperini:1991ak}.  Consider the leading order theory describing the universal massless sector of closed strings in $d+1$ dimensions. This corresponds to the well known Neveu-Schwarz action depending on the string frame metric, $\mathbb{g}_{\mu\nu}$, the antisymmetric Kalb-Ramond field, $\mathbb{b}_{\mu\nu}$, and the scalar dilaton $\phi$,
\begin{equation}\label{eq:2.1}
   I_0=\frac{1}{2\l_s^{d-1} } \int \ud^{1+d}x\sqrt{-\mathbb{g}}e^{-2\phi}\left[ R+4(\del\phi)^2 -\frac{1}{12}H^{\mu\nu\rho}H_{\mu\nu\rho}-2\Lambda \right] ,
\end{equation}
where the string length $l_s=\sqrt{\alpha'} $, $H_{\mu\nu\rho} = 3\partial_{[\mu} \mathbb{b}_{\nu\rho]}$ is the Kalb-Ramond field strength and $R$ is the Ricci scalar for the string frame metric. The cosmological term, $\Lambda$, is present for non-critical strings and scales as $l_s^{-2}$. For critical strings it vanishes, so we shall neglect it in the following. We make a cosmological ansatz for the metric, $b$-field and dilaton, 
\begin{align}\label{eq:2.2}
\mathbb{g}_{\mu\nu}&=
\begin{bmatrix}
-n^2(t) & 0\\
0 & g_{ij}(t)
\end{bmatrix},\\
 \label{eq:2.3}
\mathbb{b}_{\mu\nu}&=
\begin{bmatrix}
0 & 0\\
0 & b_{ij}(t)
\end{bmatrix},\\
\label{eq:2.4}
\phi&=\phi(t),
\end{align}
where $n(t)$ is the lapse and $g_{ij}(t)$ are the components of the spatial metric. This breaks  diffeomorphisms down to the subgroup $t \to f(t), x^i\to x^i$, with $\phi, ~g_{ij}$ and $b_{ij}$ transforming as scalars, and the lapse as a vector, $n \to n \dot f $.  In this limit, the corresponding field equations also exhibit an $\Odd$ symmetry. Indeed, the full system can be written in terms of $\Odd$ covariant objects: a generalised dilaton, 
\begin{equation}
\Phi=2\phi- \ln \sqrt{\det g},  \label{Phi}
\end{equation}
and a generalised metric 
\begin{equation}
    \mathcal{H}=
\begin{bmatrix}
g^{-1}  &  -g^{-1}b \\
bg^{-1} & g-bg^{-1}b
\end{bmatrix}.
\end{equation}
The latter transforms as a tensor, $\mathcal{H}\to\Omega^T\mathcal{H}\Omega$, under $\Omega\in \Odd$ 
where 
\begin{equation}\label{eq:2.9}
\Omega^T \eta \Omega = \eta, \qquad \eta=
\begin{bmatrix}
0 & \mathbf{1} \\
\mathbf{1}  & 0
\end{bmatrix}.
\end{equation} 
In contrast, the generalised dilaton and the lapse function are $\Odd$ scalars,  transforming trivially as $\Phi\to\Phi$ and $n\to n.$

The spatial contribution to the action can now be integrated  out, resulting in a one-dimensional, two-derivative action that can be written in a form that is manifestly invariant under the residual diffeomorphisms as well as the underlying $\Odd$ symmetry \cite{Meissner1,Gasperini:1991ak},
\begin{equation}\label{eq:2.12}
I_{0}=\kappa^2 \int \ud t \;n e^{-\Phi} \left[-(\del_n\Phi)^2 - \frac{1}{8} \tr[(\del_n S)^2] \right].
\end{equation}
where $\kappa^2>0$ is some dimensionful constant that plays no role in what follows. Note that we have introduced the following notation for the covariant time derivative
\begin{equation}\label{eq:2.11}
\del_n = \frac{1}{n(t)} \frac{\ud}{\ud t},
\end{equation}
and further defined
\begin{equation}
S= \eta \mathcal{H}
=
\begin{bmatrix}
 bg^{-1} & \;\;g-bg^{-1}b\\
g^{-1} & \;\;-g^{-1}b
\end{bmatrix}.\\
\label{eq:2.7}
\end{equation}
Note also that both $\eta$ and  $S$  are involutory matrices, squaring to the identity, $S^2=\eta^2=1$.  The action  is now seen to be manifestly invariant under $\Odd$ since $S$ transforms as\footnote{To see this, note that $S=\eta \mathcal{H} \to S'=\eta \Omega^T \mathcal{H} \Omega=\eta \Omega^T \eta S \Omega$, where we have used the fact that $\eta^2=1$. It then follows from \eqref{eq:2.9} that $\eta \Omega^T \eta=\Omega^{-1}$ and so $S'=\Omega^{-1} S \Omega$, as stated in \eqref{eq:2.8}.  In \cite{Hohm:2019ccp,Hohm:2019jgu}, these transformations are expressed in terms of $h=\Omega^{-1}$  where \eqref{eq:2.9} along with the fact that $\eta^2=1$ now implies that $h \eta h^T=\eta$. }
\begin{equation}\label{eq:2.8}
S \to S'= \Omega^{-1} S \Omega,\\
\end{equation}
We now consider the residual diffeomorphism and $\Odd$ invariant $\alpha'$ corrections to $I_0$. These terms are constructed out of $\Odd$ scalars such as $\Phi$ and its covariant  derivatives, $\del_n \Phi,\del^2_n \Phi, \ldots$ along with traces of $\Odd$ covariant terms such as $\del_nS$ and higher covariant  derivatives $\del_n^2 S, \del_n^3 S, \ldots$. Through a judicious use of equations of motion, field redefinitions, integration by parts, and useful relations of the form
\begin{align}\label{eq:2.13}
\tr(S)=\tr(\del_n S)=\tr(\del_n^2S)=...&=0 ,\\
\label{eq:2.14}
\tr((\del_n S)^{2k+1})&=0 \qquad \text{for} \qquad k=0,1,...,\\
\label{eq:2.15}
\tr(S(\del_n S)^k)&=0 \qquad \text{for} \qquad k=0,1,...,
\end{align}
we can eliminate all higher order derivatives in the system. Indeed, Holm and Zwiebach \cite{Hohm:2019jgu} showed that the most general action at all order in $\alpha'$ must take the following form
\begin{equation} \label{HZact}
I=\kappa^2 \int dt \ n e^{-\Phi}\left[-(\del_n \Phi)^2+\sum_{k=1}^\infty c_k \alpha'^{k-1} \tr[(\del_n S)^{2k}] +\textrm{multitrace} \right]
\end{equation}
where $c_1=-\frac18$ to match the leading order Neveu-Schwarz action \eqref{eq:2.12}. For any particular string theory, there ought to be a tower of dimensionless coefficients of higher order operators given by $c_k$, along with a tower of coefficients for the multitrace operators.

 \section{Isotropic cosmologies} \label{sec:iso}
In the case of isotropic and spatially flat cosmology, $g_{ij}=a^2(t)\delta_{ij}$, with a vanishing $b$ field, the system simplifies considerably and the action \eqref{HZact}  can be written in the following compact form \cite{Hohm:2019jgu}:
\begin{equation}\label{eq:2.20}
I = \kappa^2\int \ud t \;n e^{-\Phi} \left[ - \Phi^2_n + \frac{1}{\alpha'}f(\sqrt{\alpha'}H) \right],
\end{equation}
where we have introduced the shorthand $\Phi_n=\del_n \Phi, \Phi_{nn}=\del_n^2 \Phi, \ldots$ and define the string frame Hubble factor as $H=\del_n \ln a$. 
Matching with \eqref{HZact} we note that the function $f$ is analytic in a neigbourhood of the origin where it takes the following form
\begin{align}\label{eq:2.21}
f(\sqrt{\alpha' } H) &= \sum^{\infty}_{k=1} d_k \alpha'^{k}  H^{2k}, \qquad d_1={ d}.
\end{align}
and the dimensionless coefficients  $d_k$ can, in principle,  be determined from the $c_k$ and the coefficients of the multitrace operators. Note that $f$ is a manifestly even function in $H$. This follows from the $\Odd$ symmetry which takes the form of a simple scale factor duality, $a \to 1/a$ and so $H \to -H$, in the homogeneous and isotropic limit. 

In the remainder of the paper, we choose units where $\alpha'=1$. It will also be convenient to represent the scale factor through its logarithm,  $q=\ln a$.   The equations of motion for the full system  can be obtained from varying the action with respect to the lapse, the dilaton and the scale factor. This yields the equations $E_a=0$ where $a=n, \Phi, q$ and 
\begin{align}\label{eq:3.3}
E_n&=\frac{\delta I}{\delta n} =\kappa^2  e^{-\Phi} \left[\Phi^2_n - L_f(H)\right],\\
\label{eq:3.4}
E_{\Phi}&=\frac{\delta I}{\delta \Phi} =\kappa^2 n e^{-\Phi} \left[2\Phi_{nn} - \Phi^2_n- f(H) \right],\\
\label{eq:3.5}
E_{q}&=\frac{\delta I}{\delta q} =-\kappa^2 \frac{\ud}{\ud t}\left[e^{-\Phi} f'(H)\right].
\end{align}
Here, $
    L_f(H) =Hf'(H)-f(H)
$
denotes the Legendre transform of $f$ and $f'(H)=\ud f/\ud H$. Thanks to the residual diffeomorphisms, these equations are not independent and satisfy the following identity
\begin{equation}
\frac{\ud}{\ud t} E_n=\Phi_n E_\Phi+H E_q.
\end{equation}
We can include matter in an $\Odd$ invariant way by including a matter action of the form $S_m[\Phi, \mathcal{S}, \chi_I]$, where $\chi_I(t)$ are $\Odd$ invariant matter fields \cite{Gasperini:1991ak,Bernardo:2019bkz}. For a homogeneous and isotropic cosmology, this is equivalent to $S_m[\Phi, q, \chi_I]$. The vacuum equations of motion are now sourced as follows
\begin{equation}
    E_n=\bar \rho, \quad E_q=-d n \bar p,  \quad  E_\Phi=\frac12 n \bar \sigma
\end{equation}
where the $\Odd$ covariant energy density, pressure and dilaton charge are respectively given by 
\begin{equation}
    \bar \rho=- \frac{\delta S_m}{\delta n}\Big|_{\Phi, q},\quad  \bar p=\frac{1}{dn} \frac{\delta S_m}{\delta q}\Big|_{n, \Phi}, \quad \bar \sigma=-\frac{2}{n} \frac{\delta S_m}{\delta \Phi}\Big|_{n, q}.
\end{equation}
Note that the $\Odd$ covariant sources differ from their diffeomorphism invariant counterparts by a factor of $\sqrt{-g}=a^d$, namely, $\rho=\bar \rho/a^d, ~p=\bar p/a^d, ~\sigma=\bar \sigma/a^d$. For the pressure, we should also note that $p$ is now defined by holding $\Phi=2\phi-d q$ fixed when we vary the matter action. This differs from the usual definition of pressure given by  $$p_\text{standard}=\frac{1}{dn a^d} \frac{\delta S_m}{\delta q}\Big|_{n, \phi},$$  where $\phi$ is held fixed instead. It is easy enough to show that  the two pressures can be related using the dilaton charge, $p_\text{standard}= p+\frac12  \sigma$ \cite{Quintin:2021eup}. Note also that diffeomorphism invariance in the matter sector implies that 
\begin{equation}
    \del_n \rho+d H (\rho+p)=\frac12  \Phi_n \sigma.
\end{equation}
Although we have included the 
matter sources for completeness, in the remainder of this paper we shall mostly neglect them, since our focus will be on finding consistent solutions to the vacuum field equation. In particular, we will seek conditions for consistent de Sitter solutions for the metric in  both the string and the Einstein frame.  We  refer the reader to \cite{Gasperini:1991ak,Bernardo:2019bkz} for more details on matter couplings. 

\subsection{dS solutions in the string frame} \label{sec:SF}
It was already shown in \cite{Hohm:2019ccp,Hohm:2019jgu} that string frame de Sitter solutions are possible in this class of $\alpha'$ complete cosmologies   for suitable choices of the function $f(H)$. These solutions are necessarily non-perturbative in $\alpha'$, and could not have been found using standard techniques in perturbative string theory.  However, it has also been  shown that these solutions are unstable under vacuum fluctuations \cite{Bernardo:2020zlc,Bieniek:2022mrv,Bernardo:2022nex}. We will review  the derivation of these solutions and add some additional insight into the nature of the instability. 

We know that in a neighbourhood of the origin, the function $f(H)$ is even and  admits the following Taylor expansion in $H^2$
\begin{equation} \label{forigin}
    f(H)={d}H^2 +d_2H^4 +\ldots
\end{equation}
Although we cannot assume that the function remains analytic for all values of $H$, we shall assume that it is twice differentiable in order to have well defined field equations. We shall also assume it to be an even function in order to preserve the $\Odd$ symmetry.

Here we are interested in the properties of the function in a neighbourhood of a string frame de Sitter solution with $H=H_c>0$, for some constant $H_c$.  Whilst it is tempting to  motivate this with the current phase of accelerated expansion, it is important to note that we  are working in $d+1$ dimensions.  For the critical dimension ($d+1=10, 26$), we must perform  compactification   down to four dimensions before making contact with observation. Therefore, we remain agnostic as to the value of $H_c$.

Setting $E_a=0$ for $a=n, \Phi, q$, we  see  that $H=H_c>0$, for some constant $H_c$, if and only if \cite{Rost}
\begin{equation}
    \Phi^2_n=-f(H_c)=Q^2,  \quad  f'(H_c)=0,
\end{equation}
for some real constant $Q$. This means that $f(H)$ must have a minimum at $H=H_c$. Thanks to the $\Odd$ symmetry, it must also have a minimum at $H=-H_c$. This suggests  the following general form
\begin{equation} \label{fsds}
    f(H)=-Q^2+ \frac{g(H)}{H_c^4} (H^2-H_c^2)^2 
\end{equation}
for some even function $g(H)$ that admits a Taylor expansion in $H^2-H_c^2$
\begin{equation}
    g(H)=g_0+g_1 (H^2-H_c^2)+ \ldots.
\end{equation}
For the form of $f(H)$ given in \eqref{fsds} to overlap with the form  close to the origin  \eqref{forigin}, we obtain the following constraints
\begin{equation}
  g(0)=Q^2, \qquad g'(0)=0, \qquad g''(0)=2{ d}+\frac{4Q^2}{H_c^2}.  \label{bcs}
\end{equation}
It is also important to establish the stability of these de Sitter solutions. Of course, a full analysis of cosmological perturbations requires knowledge of the underlying theory beyond the homogeneous limit under consideration. Although there has been some interesting recent work in this direction \cite{Codina:2023fhy}, it is well known that $\Odd$ invariance in $d+1$ dimensions is at odds (!) with manifest diffeomorphism invariance, as evidenced through double field theory \cite{Hohm:2013jaa,Hohm:2014xsa} (see also \cite{Brustein:1998kq} for a related duality observed in cosmological perturbations).  Faced with these limitations, we can only consider the stability of solutions under homogeneous perturbations captured by the general action \eqref{HZact}. Instabilities for homogeneous and isotropic perturbations of string frame de Sitter solutions were identified in \cite{Bernardo:2020zlc,Rost,Bieniek:2022mrv,Bernardo:2022nex} with subsequent extensions to anisotropic perturbations \cite{Bieniek:2023ubx}.  Here we take a subtlely different approach to those works but draw the same conclusions.

We consider only homogeneous and isotropic perturbations, so that the system of equations $E_i=0$ remains unchanged. Assuming $f''(H) \neq 0$, these can be written in the following form
\begin{equation}
 f'(H)=c  e^\Phi, \qquad  \del_n H=s \frac{\sqrt{L_f(H)} f'(H)}{f''(H) }, \label{vaceqns}
\end{equation}
where $c$ is a constant of integration. Here $s=\pm1$ corresponds to the sign of $\dot \Phi=d\Phi/dt$, appearing when we take the square root in \eqref{eq:3.3}. This ability to take either sign was identified as a potential source of error in the stability analysis, as discussed in \cite{Bernardo:2022nex}. Next we will show how the choice of sign plays a crucial role in opening up the unstable channel.

We  begin by making a choice of the time co-ordinate by setting the lapse function to unity, $n=1$, and solve for the Hubble parameter in a neighbourhood of the vacuum, giving 
\begin{equation}
   H =H_c+\delta H
\end{equation}
where $\delta H$ satisfies
\begin{equation}
    \dot {\delta H} \approx \begin{cases} s  \sqrt{ H_c f''(H_c)}  \delta H^{3/2} & Q=0 \\
    s |Q| \delta H & Q \neq 0
    \end{cases}
\end{equation}
For $Q=0$, this gives the consistent solution
\begin{equation}
    \delta H \approx  \frac{4 }{H_c f''(H_c)(t-t_c)^2}
\end{equation}
with $s=-\text{sgn}(t-t_c)$. We immediately see the unstable channel if and only if the initial time $t_i <t_c$, or equivalently $s=+1$.   For $Q \neq 0$, we also get an instability for $s=+1$, as
\begin{equation}
    \delta H \approx 
 e^{s|Q|(t-t_c)}    
\end{equation}
In all cases, the  instability is excited if and only if $\dot \Phi(t_i)>0$. These results are consistent with the instability analysis presented in \cite{Rost}.

\subsection{dS solutions in Einstein frame} \label{sec:EF}
We now turn our attention to the Einstein frame. Einstein frame de Sitter solutions with constant $\dot \Phi$ were obtained in \cite{Bernardo:2019bkz} in the presence of a source with $\bar p=-\bar \rho$. Here we shall present an Einstein frame de Sitter metric as solutions to the {\it vacuum} equations of motion for bespoke choices of the function $f(H)$.   Although the solutions are stable, we will also show that the required form of $f(H)$ cannot be smoothly connected to the leading order Neveu-Schwarz theory in a neighbourhood of $H=0$. This agrees with the conclusions drawn in \cite{EFdS} and \cite{Rost}. 

The string frame and the Einstein frame are related by a conformal transformation.  Labelling the Einstein frame with the letter ``E", we write $\mathbb{g}_{\mu\nu} =\mathcal{A}^2 \mathbb{g}_{\mu\nu}^E$  and so $n = \mathcal{A} n_E $, $a_S = \mathcal{A} a_E$, where $\mathcal{A}$ is the conformal factor to be derived presently. Focussing on the leading order curvature term in the action, we note that
\begin{equation}
\int \ud^{1+d}x\sqrt{-\mathbb{g}}e^{-2\phi} R=\int \ud^{1+d}x\sqrt{-\mathbb{g}_E}e^{-2\phi}\mathcal{A}^{d-1} \left[R_E +\ldots\right]
\end{equation}
where the ellipsis include derivatives of $\mathcal{A}$ and $R_E$ is the Ricci scalar for the transformed metric. For the latter to coincide with the Einstein frame metric, we must choose $\mathcal{A}=e^\frac{2\phi}{d-1}$.

Setting  $E_i=0$ for $i=n, \Phi, q$, we see that the vacuum field equations are equivalent to
\begin{equation}
   f'(H)=c e^\Phi, \qquad \Phi_n=s\sqrt{L_f(H)}  \label{vaceqnsE}
\end{equation}
where we recall that $c$ is a constant of integration and $s=\pm 1$. We wish to recast these equations in terms of the Hubble parameter defined in the Einstein frame
\begin{equation}
    H_E=\del_{n_E} q_E
\end{equation}
where $\del_{n_E}=\frac{1}{n_E}\frac{d}{dt}$ and  $q_E=\ln a_E$. From the definition of the generalised dilaton 
$\Phi=2\phi -dq$ and the conformal mapping  between frames $q-q_E=\ln(n/n_E)=\frac{2\phi}{d-1}$, we can derive the following useful relations
\begin{equation}
  \Phi=-q-(d-1) q_E  \implies \Phi_n=-H-(d-1)\frac{n_E}{n} H_E \label{map1}
\end{equation}
and
\begin{equation}
    \phi=-\frac{(d-1)}{2} (\Phi+dq_E) \implies  \frac{n}{n_E}=\frac{e^{-\Phi}}{\V} \label{map2}
\end{equation}
where $\V=a_E^d$ is volume scale factor in Einstein frame. 

As  first noted in \cite{Rost}, we can inherit an Einstein frame de Sitter vacuum from the string frame de Sitter vacua presented in the previous section, by fixing the dilaton, $\phi$, to be constant. This corresponds to the case where $\Phi_n=-dH_c$, with $H_c$ the constant Hubble scale in string frame. These solutions preserve all the de Sitter isometries and would seem to be in contradiction of the classical no-go theorem of \cite{Sethi}.  This suggests that the corresponding form of $f(H)$ does not emerge from classical heterotic or type II strings. For this reason, we will focus instead on quasi-de Sitter solutions in Einstein frame, where the dilaton breaks the de Sitter isometries.

After some manipulations,  the system of equations \eqref{vaceqnsE}, \eqref{map1} and \eqref{map2} can be expressed as follows
\begin{equation} \label{sys}
e^\Phi=\frac{f'(H)}{c}, \qquad \frac{n}{n_E}=\frac{c }{f'(H)\V}, \qquad \partial_{n_E} \V=c \mathcal{F}(H), \qquad  \partial_{n_E} H=\frac{c \mathcal{G}(H)}{\V}
\end{equation}
where 
\begin{equation}
\mathcal{F}(H)=\frac{d}{1-d}\left[\frac{H+s \sqrt{L}}{f'(H)}\right], \qquad \mathcal{G}(H)=\frac{s \sqrt{L}}{f''(H)}
\end{equation}
and we have assumed $f'(H)f''(H)\neq 0$. Note that the choice $f'(H) \neq 0$ is consistent with a time dependent dilaton. To explore the possibility of a de Sitter solutions in the Einstein frame, it is convenient to construct 
\begin{equation}
\del_{n_E} H_E =dH_E^2 \left[\frac{\V \del_{n_E}^2 \V}{(\del_{n_E} \V)^2} -1\right]=dH_E^2\left[\frac{\mathcal{F}'(H) \mathcal{G}(H)}{\mathcal{F}(H)^2}-1 \right],\qquad c \mathcal{F}(H)\neq 0. \label{dnHE}
\end{equation}
If we choose the function $f(H)$ such that the right hand of this equation vanishes identically, the Hubble parameter in Einstein frame is guaranteed to be constant, $H_E=H_\star$, for a solution to the vacuum field equations.  Thus, the existence of the de Sitter solution in Einstein frame requires that
\begin{equation} \label{cond}
\left(\frac{1}{\mathcal{F}}\right)'+\frac{1}{\mathcal{G}}=0,
\end{equation}
or in other words, that $f(H)$ satisfies the following differential equation, 
\begin{equation}\label{eq:4.21}
\frac{d+1}{d-1} s H \sqrt{H f'-f} f'' + \frac{1}{d-1}(H f'-f)f'' + \frac{d}{d-1} H^2 f'' + s \sqrt{H f'-f} f' + \frac{1}{2}H f' f''=0.
\end{equation}
 Note further that we can use the third equation in  \eqref{sys}  to show that the Hubble parameters in the Einstein and string frame are related according to the following formula 
\begin{equation} \label{hst}
H_E =\frac{ c \mathcal{F}(H)}{d\V}.
\end{equation}
For a suitable choice of $f(H)$ satisfying \eqref{eq:4.21}, the corresponding constant curvature solution is stable under vacuum perturbations, as it is the {\it only} solution to \eqref{dnHE}. 

Finding a suitable choice of $f(H)$ is not easy as the differential equation \eqref{eq:4.21} is challenging to solve in full generality.  However, if we assume that $H>0$, it does admit the following analytic solution when $s=-1$, 
\begin{equation}\label{eq:4.23}
\bar f(H) = (H + H_m)^2 - H^2_m,
\end{equation}
where $H_m$ is a constant \cite{Rost}. 

With  $f=\bar f$ and taking the negative ($s=-1$) root  we find that $\mathcal{F}=0$ and $\mathcal{G}=-H/2$.  The vanishing of $\mathcal{F}$ suggests this solution lies outside of the original regime of validility for which \eqref{eq:4.21} was derived.  Nevertheless, it is interesting to explore the solutions to the system of equations \eqref{sys} in this particular case. It turns out that they imply a constant but vanishing Hubble rate in Einstein frame
\begin{equation}
H_\star=0, 
\end{equation}
consistent with \eqref{hst}, along with the following solutions for the string frame Hubble parameter, the dilaton and the lapse, 
\begin{eqnarray}
H(t) &=& H_c \exp\left(-\frac{ct}{2a_0^d}\right), \label{hsol}\\
\Phi(t) &=& \ln\left[\frac{2}{c}\left(H_c \exp\left(-\frac{ct}{2a_0^d}\right)+H_m\right) \right],  \\
n(t) &=& \frac{c}{2 a_0^d}\left[H_c\exp\left(-\frac{ct}{2a_0^d}\right)+H_m\right]^{-1}, \label{nsol}
\end{eqnarray}
where $H_c$ is a dimensionful constant of integration and we have assumed $d \neq 1$ and $c\neq 0$. Clearly the solution runs into a singularity as $H \to -H_m$  although this can be trivially avoided by taking $H_m>0$.

\begin{figure}[h]
    \centering
    \includegraphics[width=.45\textwidth]{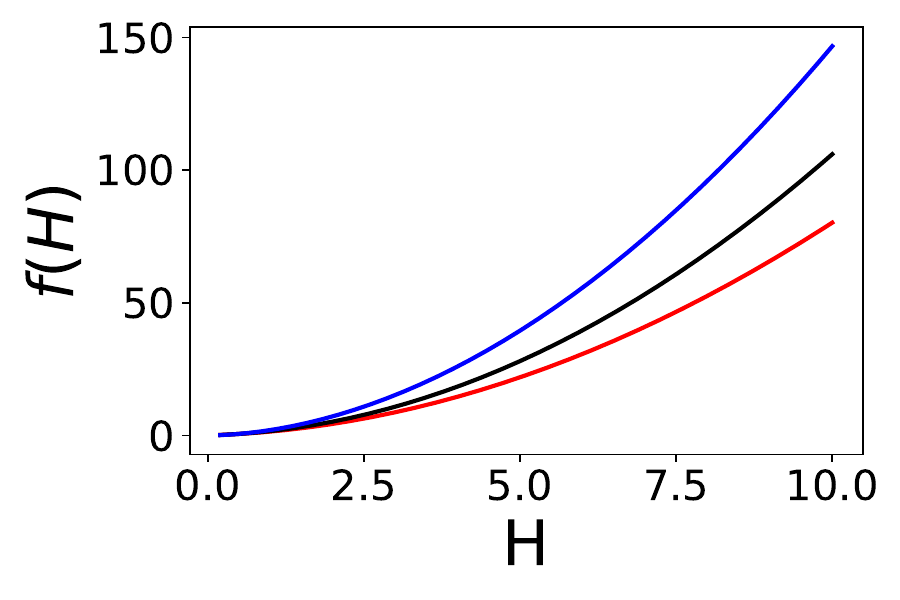}
    \caption{Examples of numerically computed $f(H)$, with the central black curve corresponding to $\bar f(H)$.}
    \label{fig:f_H}
\end{figure}

\begin{figure}[h]
    \centering
    \includegraphics[width=.45\textwidth]{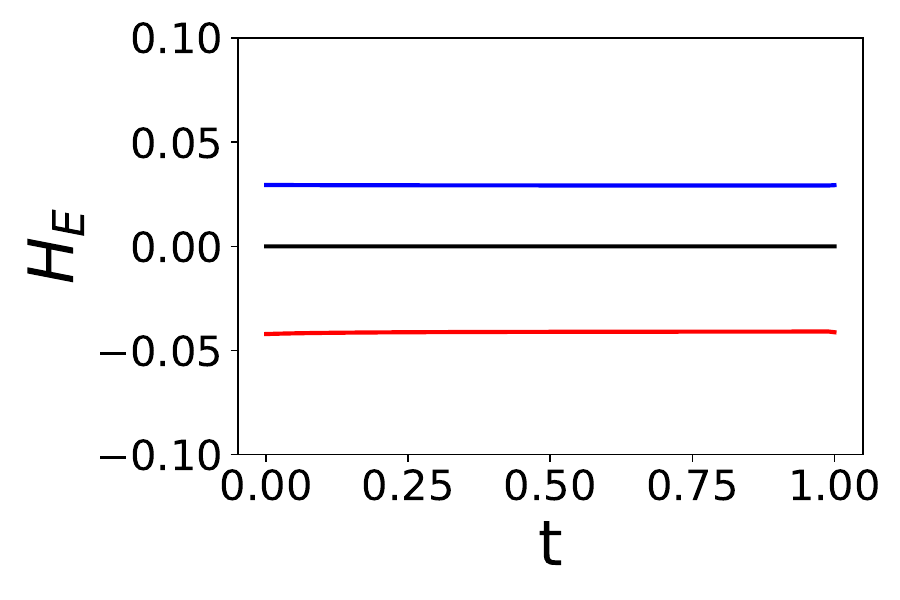}
    \caption{Time evolution of the Einstein-frame Hubble parameter for the $f(H)$ given in Fig. \ref{fig:f_H}}. 
    \label{fig:HE}
\end{figure}

We have not been able to find an exact  solution to \eqref{eq:4.21} that gives non-vanishing $\mathcal{F}$ and, correspondingly,  $H_\star>0$. However, numerical solutions are presented in \figref{fig:f_H}, with the corresponding value of $H_E$ shown in \figref{fig:HE}, where we recall that we have set units such that $\alpha'=1$.

We can also seek a perturbative solution to \eqref{eq:4.21}. To this end, we consider  
\begin{equation}\label{eq:4.24}
f(H) = \bar f(H) \left[1+ \epsilon (H)\right],
\end{equation}
where we assume $|\epsilon|\ll1$. Plugging this into \eqref{eq:4.21} (with $s=-1$) and keeping terms up to linear order in $\epsilon$, we arrive at the linear differential equation
\begin{equation}
(H_m-H)(2H_m-H)H \epsilon''(H)+(2H_m^2-3H H_m+2 H^2) \epsilon'(H)-H_m \epsilon (H)=0
\end{equation}
with the following solution 
\begin{equation}\label{eq:4.30}
\epsilon (H) =  \frac{C_1 H_m + C_2 H + h_{m} C_2 \ln \left|\frac{H}{H_m}\right|}{H+2h_{m}},
\end{equation}
where $C_1$ and $C_2$ are constants of integration. Note that $C_1$ can be removed by redefining $H_m \to H_m\left(1-C_1/2\right)$ and working to linear order in $C_1$. Thus we can set $C_1=0$ without loss of generality and express \eqref{eq:4.30} as a simple  function of the ratio $\delta=H/H_m$, 
\begin{equation} \label{eps}
\epsilon (H)=C_2\left[\frac{\delta+\ln |\delta|}{\delta+2 }\right].
\end{equation}
Recall that this expression is valid as long as $|\epsilon(H)| \ll 1$. If we assume  $|C_2|\ll 1 $ then this condition is violated if and only if $\delta$  strays too close to the singularities at $\delta=0, -2$. More precisely, we have $\epsilon(H) \ll 1$ whenever $|C_2|\ll 1 $, provided
$$
|\delta|  \gg e^{-2/|C_2|}\ , \quad \Big| \delta+2 \Big| \gg |C_2| \left(2 -\ln 2 \right).
$$
We now revisit the field equations \eqref{sys} and \eqref{cond} using the perturbed form of $f$ given by \eqref{eps}. To leading order in $|C_2| \ll 1$ \eqref{cond} gives the following expression for Hubble rate in Einstein frame
\begin{equation}
H_E \approx \frac{C_2 c}{4(d-1)\V} \label{HEeq}.
\end{equation}
If we assume $H_E = H_\star$ for some constant $H_\star$ and  set $n_E=1$, it follows that that we have $\V=a_0^d e^{d H_\star t}$. We can then solve \eqref{HEeq} explicitly, giving
\begin{equation}
    H_{\star} \approx \frac{1}{dt} W\left(\frac{dC_2 ct}{4(d-1)a_0^d}\right) \approx \frac{C_2 c}{4(d-1)a_0^d}
\end{equation}
where we have used the fact that the Lambert W function $W(x) \approx x$ for small $x$.  We see that we have a constant and {\it non}-vanishing  Hubble rate in the Einstein frame, with the approximation valid as long as $|t| \ll 4(d-1)a_0^d/d|c||C_2|$. Since $H_\star$ is required to be positive in an expanding universe, we assume that $C_2 c>0$. The remaining fields are dominated by their non-vanishing zeroth order (in $\epsilon$) solutions \eqref{hsol}-\eqref{nsol}. 
Recall that in a neighbourhood of the origin, perturbative string theory suggests that the function $f(H)$ should admit a Taylor expansion of the form \eqref{forigin}.  We will now show that any solution to the differential equation \eqref{eq:4.21} valid for $H>0$ cannot be analytically continued towards a solution of the form \eqref{forigin} in a neighbourhood of $H=0$.

To see this, we begin by assuming that $f(H)$ admits a Taylor expansions of the form \eqref{forigin} in a neighbourhood of the origin 
\begin{equation}
    f(H)={ d} H^2+\mathcal{O}(H^4) \label{forigin2}.
\end{equation}
In this neighbourhood, it immediately follows that
\begin{equation}
   \frac{\mathcal{F}'(H) \mathcal{G}(H)}{\mathcal{F}(H)^2}=\mathcal{O}(H^2).
\end{equation}
In order to satisfy \eqref{eq:4.21}, the right hand side of this expression should be equal to unity, which is clearly not possible in a neighbourhood of $H=0$, regardless of the sign of $s=\pm 1$.  Thus, it is not possible to smoothly connect a solution to \eqref{eq:4.21} to the desired leading order Neveu-Schwarz theory at low curvatures. However, perhaps this is too demanding. As noted earlier, for the field equations to be well defined, we only need the form of $f(H)$ to be twice differentiable. For any finite $N$, it is certainly possible to match the first $N$ derivatives of a solution to \eqref{eq:4.21} to a function of the form \eqref{forigin2}, at some  value of $H=H_\text{match}>0$.

It is, perhaps, optimistic to expect such a matching  to emerge from a non-perturbative understanding of string theory. Nevertheless, we could imagine a discontinuity in the higher order derivatives of $f(H)$ to be a remnant  of the truncation to the massless sector of the theory. Indeed, if it emerges from string oscillators with string scale masses, we might expect $H_\text{match} \sim 1$ (in units of $\alpha'=1$).

Even if this desired form of $f(H)$ can be justified, it is clear that we can only trust our de Sitter solution in the regime where $H>H_\text{match}$. However, recall that $H$ is the Hubble parameter in the {\it string frame} and this will change over time. If the system evolves into a  regime where $H<H_\text{match}$, the form of $f(H)$ will no longer admit de Sitter solutions and the Einstein frame de Sitter metric is destabilised.  

To avoid this instability, we need to stop $H$ evolving into the unstable region with $H<H_\text{match}$ in either time direction.  In other words, we need $H$ to have a turning point in its evolution at some $H_\text{turn}>H_\text{match}$. At such a turning point, the last equation in \eqref{sys} tells us that $\mathcal{G}(H_\text{turn})=0$. Since $H_\text{turn}>H_\text{match}$ by assumption, the  differential equation \eqref{cond} ought to be satisfied at $H=H_\text{turn}$, further implying that $\mathcal{F}(H_\text{turn})=0$ must have a zero  at the same point.  Since we assume that $f(H)$ is twice differentiable, $\mathcal{F}(H_\text{turn})=\mathcal{G}(H_\text{turn})=0$ if and only if $H_\text{turn}=0$. This contradicts the underlying assumption that $H_\text{turn}>H_\text{match}>0$. It follows that  our de Sitter solution in Einstein frame cannot be eternal. 

\section{Anisotropic cosmologies} \label{sec:aniso}
We now turn our attention to anisotropic cosmologies in $d+1$  spacetime and consider a string frame line element of the form
\begin{equation}
    ds^2=-n(t)^2 dt^2 +\sum_{i=1}^d a_i(t)^2 dx_i^2,
\end{equation}
where $a_i(t)$ is the scale factor along the $i$th spatial direction. A detailed derivation  of the $\Odd$ invariant action requires a careful computation of all the terms in \eqref{HZact}, including the multitrace operators. Instead of computing that explicitly, we note that the action will take the form 
\begin{equation}\label{action-aniso}
I = \kappa^2\int \ud t \;n e^{-\Phi} \left[ - \Phi^2_n + f(H_i) \right],
\end{equation}
where we have assumed units with $\alpha'=1$. The Hubble parameter along the $i$th direction is given by $H_i=\del_n q_i$ for $q_i=\ln a_i$, while the generalised dilaton is given in terms of the standard dilaton via the following relation
\begin{equation}
    \Phi=2\phi-\sum_i q_i.
\end{equation}
To match with the leading order Neveu-Schwarz action \eqref{eq:2.1} at low curvatures, we expect that $f$ admits a Taylor expansion near the origin such that
\begin{equation} \label{NS-aniso}
    f(H_i) =\sum_i H_i^2 + \mathcal{O}(H^4).
\end{equation}
The $\Odd$ symmetry now includes scale factor duality along each direction, $a_i \to 1/a_i$ and scale factor exchange $a_i \leftrightarrow a_j$ for each $i$ and $j$. This suggests that the function $f(H_i)$ can also be expressed as
\begin{equation}
    f(H_i)=F\left(\hat e^d_i(H_j^2)\right) \label{elemsym},
\end{equation}
where we define the normalised elementary symmetric polynomials 
\begin{equation}
    \hat e_i^d(X_j)=\frac{1}{\binom{d}{i}} \sum_{1 \leq j_1 < \ldots j_i \leq d} X_{j_1} \ldots X_{j_i}
\end{equation}
for $i=1, \ldots d$.

Varying the action \eqref{action-aniso} with respect to the lapse, the dilaton and each of the scale factors, we obtain $E_a=0$ for $a=n, \Phi, q_i$, where
\begin{align}\label{En-aniso}
E_n&=\frac{\delta I}{\delta n} =\kappa^2  e^{-\Phi} \left[\Phi^2_n - L_f(H_i)\right],\\
\label{Ephi-aniso}
E_{\Phi}&=\frac{\delta I}{\delta \Phi} =\kappa^2 n e^{-\Phi} \left[2\Phi_{nn} - \Phi^2_n- f(H_i) \right],\\
\label{Ei-aniso}
E_{q_i}&=\frac{\delta I}{\delta q_i} =-\kappa^2 \frac{\ud}{\ud t}\left[e^{-\Phi} f_i\right],
\end{align}
where we denote partial derivatives $f_i=\del f/\del H_i$ and the Legendre transform $L_f(H_i) =\sum_i H_i f_i-f(H_i)$. It is also convenient to define the average Hubble parameter in string frame as $H=\frac{1}{d} \sum_i H_i$.

\subsection{dS solutions in the string frame} \label{sec:SFan}
We now seek solutions to the anisotropic field equations corresponding to the string frame de Sitter vacua. 
Using similar arguments to the isotropic case in section \ref{sec:SF}, we see that  there is an isotropic solution to \eqref{En-aniso} to \eqref{Ei-aniso} with $H_i=H_c>0$ for all $i$ and some constant $H_c$ if and only if
\begin{equation}
    \Phi_n^2=-f(H_c, \ldots , H_c)=Q^2, \qquad f_i(H_c, \ldots , H_c)=0.
\end{equation}
whereb $Q$ is a constant. Given the form of $f$ dictated by the $\Odd$ symmetry \eqref{elemsym}, this suggests that
\begin{equation}
    f(H_i)=-Q^2+\sum_j \frac{g_j(H_l)}{H_c^{4j}}(\hat e_j^d(H_l^2)-H_c^{2j})^2
\end{equation}
where  the functions $g_j(H_k)$ take the $\Odd$ symmetric form \eqref{elemsym}  and admit a Taylor expansion in each of $\hat e_j^d(H_l^2)-H_c^{2j}$,
\begin{equation}
    g_j(H_k)=g_{j0}+\sum_k g_{jk}(\hat e_k^d(H_l^2)-H_c^{2k})+\ldots.
\end{equation}
Close to the origin, the form of $f$ should coincide with \eqref{NS-aniso} to recover the leading order Neveu-Schwarz theory at small curvatures.
% This implies that
% \begin{equation}
%     \sum_j g_j(0, \ldots, 0)=Q^2, \qquad \sum_j \del_{H_k} g_j(0, \ldots, 0)=0, \qquad \sum_j \del_{H_k}  \del_{H_l }g_j(0, \ldots, 0)=2\delta_{kl}. \label{bcsan}
% \end{equation}
On the isotropic line where  $H_i=H$ for all $i$,  the form of $f$ is compatible with the isotropic solutions \eqref{fsds}, as expected
\begin{equation}
    f(H_i)|_\text{iso}=-Q^2+\sum_j \frac{g_j(H,\ldots, H)}{H_c^{4j}}(H^{2j}-H_c^{2j})^2.
\end{equation}
It immediately follows that the corresponding de Sitter solution will once again be unstable to isotropic fluctuations when $\Phi_n(t_i)>0$. 

\subsection{dS solutions in the Einstein frame}
\label{sec:EFan}
We now switch our attention to the Einstein frame to see if the family of de Sitter solutions found in the isotropic framework \cite{EFdS} carries over to the anisotropic framework. As usual, we perform a conformal transformation to switch to Einstein frame, with  $n=\mathcal{A} n_E$ and $a_i=\mathcal{A} a_i^E$  where $\mathcal{A}=e^\frac{2 \phi}{d-1}$. We therefore  have $\ln (n/n_E)=q_i-q_i^E=\frac{2\phi}{d-1}$ for $q_i^E=\ln a_i^E$. This allows us to extract the following useful relation
\begin{equation}
    \phi=-\frac{d-1}{2} \left(\Phi+\sum_i q_i^E\right),
\end{equation}
implying 
\begin{equation}
    \frac{n}{n_E}=\frac{e^{-\Phi}}{\V}, \qquad q_i-q_i^E=-\Phi-\ln \V
\end{equation}
where $\V=\prod_i (a_i^{E})^{k_i}$ is the Einstein frame volume scale factor. 

In the Einstein frame, we define Hubble parameters $H_i^E=\del_{n_E} q_i^E$ and the corresponding average $H^E=\frac{1}{d}\sum_i H_i^E$. These are easily related to their string frame counterparts via the relation
\begin{equation}
    H_i=\frac{n_E}{n} \left(H_i^E-dH^E\right)-\Phi_n.
\end{equation}

From the field equations \eqref{En-aniso} to \eqref{Ei-aniso} we can derive expressions for the generalised velocities in Einstein frame
\begin{equation} \label{aniso-sys2}
  \del_{n_E} \Phi= \frac{e^{-\Phi}}{\V}s\sqrt{L_f} , 
  \qquad H_i^E-H^E= \frac{e^{-\Phi}}{\V} \left(H_i-H\right),
  \qquad H^E= \frac{e^{-\Phi}}{\V} \frac{\Sigma}{d} \mathcal{F}
  \end{equation}
  where 
\begin{equation}
\mathcal{F}=\frac{d}{1-d} \left(\frac{H+s\sqrt{L_f}}{\Sigma}\right), \qquad \Sigma=\sqrt{\sum_j f_j^2}
\end{equation}
and $s=\pm 1$, depending on the sign of $\Phi_n$. 
  From equation \eqref{Ei-aniso}, we have that 
 \begin{equation}
e^{-\Phi} f_i=c_i \implies e^{-\Phi} \Sigma=|\underline c|\equiv\sqrt{\sum_j c_j^2}
    \end{equation} 
for some constants $c_i$.  We can now derive the following constraints on the dynamics of the system, analogous to the last two equations in \eqref{sys}
  \begin{equation}
  \qquad   \del_{n_E} \V=| \underline c|\mathcal{F}, \qquad \del_{n_E} H_i=\frac{| \underline c|\mathcal{G}_i}{\V} 
\end{equation} 
where 
\begin{equation}
     \mathcal{G}_i=\frac{s\sqrt{L_f} \sum_j f^{ij} f_j}{\Sigma}
\end{equation}
 and $f^{ij}$ denotes the inverse of the Hessian of the function $f$, assumed to be well defined.   We proceed in a similar way to the isotropic case, and compute the rate of change of the Einstein frame Hubble parameters. We find that
\begin{equation}\label{eq:HE_i_evolution}
    \del_{n_E} H_i^E=d (H^E)^2\left[ (1+\xi_i) (\mathcal{O} \ln \mathcal{F}-1 ) +\mathcal{O}\xi_i \right],
\end{equation}
where 
\begin{align}\label{eq:O_def}
    \mathcal{O}&=\frac{1}{\mathcal{F}} \sum_j \mathcal{G}_j \del_{H_j},\\\label{eq:xi_def}
    \xi_i&=\frac{d (H_i-H)}{\Sigma \mathcal{F}}.
\end{align}
Since $\sum_j \xi_j=0$, it follows that the rate of change of the average Hubble factor is given by a somewhat simpler formula
\begin{equation}
    \del_{n_E} H^E=d (H^E)^2 (\mathcal{O} \ln \mathcal{F}-1 ), 
\end{equation}
which is strongly reminiscent of the corresponding isotropic equation \eqref{dnHE}, as expected.
We may alternatively express (\ref{eq:HE_i_evolution}) as
\begin{equation}
    \del_{n_E} H_i^E-(1+\xi_i) \del_{n_E} H^E =d (H^E)^2\mathcal{O}\xi_i.
\end{equation}
Let us now evaluate our system on the isotropic line $H_i=H$, along which we have $\xi_i=0$ from (\ref{eq:xi_def}). For consistency, it must follow that 
\begin{equation}
   (\del_{n_E} H^E)|_\text{iso}=d ( H^E)^2 (\mathcal{O} \ln \mathcal{F}-1 )|_\text{iso}, \qquad (\mathcal{O}\xi_i)|_\text{iso}=0
\end{equation}
where $|_\text{iso}$ indicates evaluation on the isotropic line.  Requiring the vacuum to be de Sitter in the Einstein frame on the isotropic line, i.e. $H_i^E=H^E=const$, now yields the following constraints
\begin{equation}
    \mathcal{O} (\ln \mathcal{F})|_\text{iso}=1,\qquad (\mathcal{O}\xi_i)|_\text{iso}=0 \label{isoeqns}.
\end{equation}
Defining   $\mathcal{G}=\sum_i\mathcal{G}_i/d$, we note that 
\begin{eqnarray}      && \mathcal{O}\left(\ln\mathcal{F}\right)=-\sum_i \mathcal{G}_i \del_{H_i} \left(\frac{1}{\mathcal{F}}\right), \\
&& \mathcal{O}(\xi_i)=\frac{1}{\Sigma \mathcal{F}^2}(\mathcal{G}_i-\mathcal{G})-\xi_i\mathcal{O}\ln (\Sigma \mathcal{F}),
\end{eqnarray}
and impose \eqref{isoeqns}, giving
\begin{equation}
  \left[\sum_i \del_{H_i} \left(\frac{1}{\mathcal{F}}\right)+\frac{1}{\mathcal{G}}\right]_\text{iso}=0, \qquad  \mathcal{G}_i|_\text{iso}=\mathcal{G}|_\text{iso} \neq 0. \label{condan}
\end{equation}
To explore the implications of these results, consider the $N$ dimensional  Euclidean space spanned by the Hubble parameters, $(H_1, \ldots, H_d)$. It is convenient to introduce a new set of coordinates $(H, v_\alpha)$, adapted to the isotropic line, defined in terms of the average  Hubble parameter $H=\sum_i  H_i/d$ and $d-1$  orthogonal directions
\begin{align}
v_\alpha&=\sum_i(H_i-H)n_i^{\alpha},
\end{align}
so that 
\begin{equation}
     H_i=H+\sum_\alpha n^{\alpha}_iv_\alpha.
\end{equation}
Here we have introduced an orthonormal set of vectors in Hubble space, $n^{\alpha}_i$,  orthogonal to the  tangent vector, $u_i=1/d$, on the isotropic line 
\begin{equation}
   \sum_i u_i n_i^{\alpha}= \frac{1}{d} \sum_i n_i^{\alpha}=0, \qquad \sum_i n_i^{\alpha} n_i^{\beta}=\delta^{\alpha \beta}.
\end{equation}
We can now think of the isotropic line as being parametrized by the average Hubble parameter, $H$, with each of the remaining coordinates vanishing there, $v_\alpha|_\text{iso}=0$.  At this point, we recall that the $\Odd$ symmetry implies that $f(H_i)$ is given in terms of the normalised elementary symmetric polynomials, as in \eqref{elemsym}. In a neighbourhood of the isotropic line, we can show that
\begin{equation}
    \hat e^d_i(H_j^2)=H^{2i} + \frac{i(d+1-2i)}{d(d-1)}H^{2(i-1)}\sum_{\alpha, \beta} \delta^{\alpha \beta} v_\alpha v_\beta+\mathcal{O}(v^3).
\end{equation}
Correspondingly we find that the function $f$ takes the form
\begin{equation}
    f(H_i)=f_0(H)+g_0(H)\sum_{\alpha, \beta} \delta^{\alpha \beta} v_\alpha v_\beta +\mathcal{O}(v^3), 
\end{equation}
where
\begin{equation}
    f_0(H)=F(H^{2}, \ldots, H^{2d}), \qquad g_0(H)=\sum_{i=1}^d  \frac{i(d+1-2i)}{d(d-1)}H^{2(i-1)} \frac{\del F}{\del \hat e^d_i }\Bigg|_{\hat e^d_j =H^{2j}}.
\end{equation}
For the de Sitter solution to exist, we require the conditions \eqref{condan} to hold. The second of these is equivalent to 
\begin{equation} \label{Geq}
    \left[ \sum  u_i \mathcal{G}_i\right]_\text{iso}=\mathcal{G}_\text{iso} \neq 0,\qquad  \left[\sum n_i^\alpha \mathcal{G}_i\right]_\text{iso}=0.
\end{equation}
Making use of the chain rule
\begin{equation}
     \frac{\del }{\del H}=\sum_i \frac{\del}{\del H_i}, \qquad \frac{\del }{\del v_\alpha }=\sum_i n_i^{(\alpha)} \frac{\del}{\del H_i},
\end{equation}
and assuming $g_0(H) \neq 0$, we immediately see that the condition \eqref{Geq} holds automatically. Meanwhile, the first condition in \eqref{condan} yields
\begin{equation}
   \left[\del_H \left(\frac{1}{\mathcal{F}}\right)+\frac{1}{\mathcal{G}}\right]_\text{iso}=0,
\end{equation}
which is equivalent  to  \eqref{eq:4.21} for $f(H)=f_0(H)$.  It follows that the Einstein frame de Sitter solutions found in the isotropic framework extend naturally to the anisotropic framework. This is because the underlying $\Odd$  symmetry ensures that deviations from isotropy only enter the theory at $\mathcal{O}(v^2) \sim \mathcal{O}((H_i-H_{j\neq i})^2)$.

\section{Conclusions} \label{sec:conc}

Can we obtain de Sitter solutions in string theory?  Although perturbative methods provide some rich phenomenology,   they have not been able to uncover de Sitter solutions that are universally accepted as consistent and fully under control. This motivates attempts to probe the non-perturbative structure of string theory. Of course, we do not have the tools to do this in full generality but progress can be made by restricting access to the Neveu-Schwarz fields in the homogeneous limit. This is because they admit an $\Odd$ symmetry that can be extended to higher order operators, extending to all orders in $\alpha'$. 

The underlying structure of these theories is easy to write down whenever we have a homogeneous and isotropic framework, where the $\Odd$ symmetry is equivalent to scale factor dulaity $a \to 1/a$. For a suitable non-linear structure,  quasi de Sitter solutions  with time dependent dilatons can be found in both the string frame and the Einstein frame. As shown previously  \cite{Bernardo:2020zlc,Rost,Bieniek:2022mrv,Bernardo:2022nex}, homogeneous and isotropic fluctuations about the string frame solutions  can trigger an instability, although we have now shown that this is excited if and only if the generalised dilaton is growing at the initial time $\dot \Phi(t_i)>0$.  The situation in the Einstein frame is even more delicate. Quasi de Sitter solutions can and do exist in the Einstein frame, for particular non-perturbative extensions of the Neveu-Schwarz theory. However, these extensions do not connect smoothly to the leading order Neveu-Schwarz theory. This means the Einstein frame de Sitter solutions cannot be eternal to both past and future. In other words, a de Sitter solution can survive for a finite period of time, or even a semi-infinite time, but at some point towards either the future or the past, the solution breaks down. The solution enters a new phase of the theory in a neighbourhood of small string frame curvature, where Einstein frame de Sitter solutions are no longer possible. Perhaps this should have been expected, as the breakdown occurs in the perturbative regime where de Sitter solutions are conjectured to be absent \cite{ds1,ds2,ds3}. 

All of these results can be extended to include Bianchi I metrics.  The $\Odd$ symmetry now plays a more interesting role than in  the isotropic case, where scale factor duality is accompanied by scale factor exchange $a_i \leftrightarrow a_j$.  This is crucial, especially for the existence of de Sitter solutions in the Einstein frame where the additional constraints required for existence in an anisotropic setting are satisfied automatically thanks to $\Odd$ symmetry. 

Although we have focused our analysis on the case of classical strings in a critical dimension, for which the constant $\Lambda=0$ in \eqref{eq:2.1}, the results are easily extended to non-critical dimensions.  This is because a non-vanishing $\Lambda \propto D-D_c$ is easily absorbed into a redefinition of the function $f$ that defines  theory to all orders in $\alpha'$, 
\begin{equation}
    f \to f-2 \Lambda.
\end{equation}
This changes the structure of the theory in neighbourhood of the origin, but little else. For the string frame de Sitter solutions in the isotropic framework, this will affect the boundary conditions \eqref{bcs}, so that we now have $g(0)=-2\Lambda$. Similar relations hold in the anisotropic case. The stability analysis is unchanged.  For the Einstein frame, we can revisit the question of whether or not the bespoke choice of $f(H)$ can be smoothly connected to the desired form in a neighbourhood of $H=0$. Focusing on the isotropic case, we set $f(H)=-2\Lambda +{ d}H^2+\mathcal{O}(H^4)$  and compute
\begin{equation}
   \frac{\mathcal{F}'(H) \mathcal{G}(H)}{\mathcal{F}(H)^2}=\frac{d-1}{d}+\mathcal{O}(H).
\end{equation}
Clearly this cannot be set equal to unity, as required for Einstein frame de Sitter solutions with a time dependent dilaton. Thus, we conclude that the situation is qualitatively unchanged from the critical case, $D=D_c$, and one cannot have eternal de Sitter solutions in the Einstein frame.

\paragraph{Acknowledgements} 
We would like to thank Heliudson Bernardo, Cliff Burgess, Guilherme Franzmann, Carmen Nu\~nez, Thomas Van Riet and Facundo Rost   for useful discussions. AP and PMS were supported by STFC consolidated grant number ST/T000732/1 and YL by an STFC studentship. RGCS was supported by a Bell-Burnell studentship.  This paper was drafted during the period of the UCU marking and assessment boycott, demonstrating work carried out by AP and PMS as employees of the University of Nottingham. For the purpose of open access, the authors have applied a CC BY public copyright licence to any Author Accepted Manuscript version arising.

\end{document}